\documentclass[showpacs,aps,prd]{revtex4}
\input epsf

\begin{document}

\title{Revisiting $D_{s0}^{*}(2317)$ as a $0^{+}$ tetraquark state from QCD sum rules}
\author{Jian-Rong Zhang}
\affiliation{Department of Physics, College of Liberal Arts and Sciences, National University of Defense Technology,
Changsha 410073, Hunan, People's Republic of China}


\begin{abstract}
Stimulated by the renewed observation of $D_{s0}^{*}(2317)$ signal and its updated mass
value $2318.3\pm1.2\pm1.2~\mbox{MeV}/c^{2}$ in the
process $e^{+}e^{-}\rightarrow D_{s}^{*+}D_{s0}^{*}(2317)^{-}+c.c.$ by BESIII Collaboration, we devote
to reinvestigate $D_{s0}^{*}(2317)$ as a $0^{+}$ tetraquark state
from QCD sum rules. Technically, four different possible currents are adopted
and high condensates up to dimension $12$ are included in the operator product expansion (OPE)
to ensure the quality of QCD sum rule analysis.
In the end,
we obtain the
mass value $2.37^{+0.50}_{-0.36}~\mbox{GeV}$ with the factorization parameter $\rho=1$ (or $2.23^{+0.78}_{-0.24}~\mbox{GeV}$
with $\rho=3$) for the
 scalar-scalar current,
which agrees well with the experimental data of $D_{s0}^{*}(2317)$ and could support
its explanation as a $0^{+}$ scalar-scalar tetraquark state.
The final result for the axial-axial configuration
is calculated to be $2.51^{+0.61}_{-0.43}~\mbox{GeV}$ with $\rho=1$ (or $2.52^{+0.76}_{-0.52}~\mbox{GeV}$ with $\rho=3$),
which is still consistent with the mass of $D_{s0}^{*}(2317)$ considering the uncertainty,
and then the possibility of $D_{s0}^{*}(2317)$ as a axial-axial tetraquark state can not be excluded.
For the pseudoscalar-pseudoscalar and the vector-vector cases, their
unsatisfactory OPE convergence makes that it is of difficulty to find
rational work windows to further acquire hadronic masses.
\end{abstract}
\maketitle

\section{Introduction}\label{sec1}
Very recently, BESIII Collaboration announced the observation of
the process $e^{+}e^{-}\rightarrow D_{s}^{*+}D_{s0}^{*}(2317)^{-}+c.c.$ for the first time
with the data sample of $567~\mbox{pb}^{-1}$
at a center-of-mass energy $\sqrt{s}=4.6~\mbox{GeV}$ \cite{BESIII}.
For the $D_{s0}^{*}(2317)$ signal, the statistical significance is reported to be $5.8\sigma$ and its mass is measured to be $2318.3\pm1.2\pm1.2~\mbox{MeV}/c^{2}$.
Historically, $D_{s0}^{*}(2317)$ was first observed by BABAR
Collaboration in the $D_{s}^{+}\pi^{0}$
invariant mass distribution \cite{BABAR,BABAR1}, which was
confirmed by CLEO Collaboration \cite{CLEO} and by Belle Collaboration \cite{Belle}.
In theory, $D_{s0}^{*}(2317)$ could be proposed as a conventional $\bar{c}s$ meson
with $J^{P}=0^{+}$ (e.g. see \cite{Colangelo}). However, one has to confront an approximate $150~\mbox{MeV}/c^{2}$ difference between
the measured mass and the theoretical results from potential model \cite{potential model} and lattice QCD \cite{lattice QCD}
calculations. In addition, the absolute branching
fraction $1.00_{-0.14}^{+0.00}\pm0.14$ for $D_{s0}^{*}(2317)^{-}\rightarrow\pi^{0}D_{s}^{-}$
newly measured by BESIII \cite{BESIII} shows
that $D_{s0}^{*}(2317)^{-}$ tends to have a significantly larger
branching fraction to $\pi^{0}D_{s}^{-}$ than to $\gamma D_{s}^{*-}$, which differs
from the expectation of the conventional $\bar{c}s$ state \cite{Godfrey}.
As a feasible scenario resolving the above discrepancy, one can suppose
$D_{s0}^{*}(2317)$ to be some multiquark system, such as a
$DK$ molecule candidate \cite{molecule}, a $\bar{c}sq\bar{q}$ tetraquark state \cite{tetraquark state}, or a mixture of
a $\bar{c}s$ meson and a tetraquark state \cite{mixture}.
In a word, it is still undetermined and even unclear for the nature of $D_{s0}^{*}(2317)$.

Especially inspired by the BESIII's new experimental result on $D_{s0}^{*}(2317)$ \cite{BESIII},
we devote to study it in the tetraquark picture, which is also
helpful to deepen one's understanding on nonperturbative
QCD.
One reliable way for evaluating
the nonperturbative effects is the QCD sum rule method \cite{svzsum}, which is an
analytic formalism firmly entrenched in QCD and
has been fruitfully applied
to many hadrons \cite{overview1,overview2,overview3,reinders,overview4}.
Concerning $D_{s0}^{*}(2317)$,
there have appeared several QCD sum rule works
to compute its mass
basing on a $\bar{c}s$ meson picture \cite{cs0,cs1,cs2,cs3,cs4,cs5,cs6,cs7}, or
taking a point of tetraquark view from QCD sum rules in
the heavy quark limit \cite{tetra0-heavy-limit} as well as
from full QCD sum rules involving condensates up to dimension $6$ or $8$
\cite{tetra1,tetra2,tetra3}.
It is known that one key point of the QCD sum rule analysis is that
both the OPE convergence and
the pole dominance should be carefully inspected.
It has already been noted that some high dimension condensates
may play an important role in some cases \cite{Zs0,Zs1,Zs2,Zs}.
To say the least,
even if high condensates
may not radically influence the OPE's character,
they are still beneficial to stabilize Borel curves.
Therefore, in order to further reveal the internal structure of $D_{s0}^{*}(2317)$,
we endeavor to perform the study of $D_{s0}^{*}(2317)$ as a $0^{+}$ tetraquark state
in QCD sum rules adopting four different possible currents and
including condensates up to dimension $12$.

The rest of the paper is organized as follows. In Sec. \ref{sec2}, $D_{s0}^{*}(2317)$ is
studied as a tetraquark state in the QCD sum
rule approach. The last part is a brief
summary.
\section{QCD sum rule study of $D_{s0}^{*}(2317)$ as a $0^{+}$ tetraquark state}\label{sec2}
\subsection{$0^{+}$ tetraquark state currents}
As one basic point of QCD sum rules, hadrons are represented by
their interpolating currents.
For a tetraquark state, its current ordinarily can be constructed as
a diquark-antidiquark configuration.
Thus, one can present following forms of $0^{+}$ tetraquark currents:
\begin{eqnarray}
j_{(I)}&=&\epsilon_{abc}\epsilon_{dec}(q_{a}^{T}C\gamma_{5}s_{b})(\bar{q}_{d}\gamma_{5}C\bar{Q}_{e}^{T})\nonumber
\end{eqnarray}
for the scalar-scalar case,
\begin{eqnarray}
j_{(II)}&=&\epsilon_{abc}\epsilon_{dec}(q_{a}^{T}Cs_{b})(\bar{q}_{d}C\bar{Q}_{e}^{T})\nonumber
\end{eqnarray}
for the pseudoscalar-pseudoscalar case,
\begin{eqnarray}
j_{(III)}&=&\epsilon_{abc}\epsilon_{dec}(q_{a}^{T}C\gamma_{\mu}s_{b})(\bar{q}_{d}\gamma^{\mu}C\bar{Q}_{e}^{T})\nonumber
\end{eqnarray}
for the axial vector-axial vector (shortened to axial-axial) case, and
\begin{eqnarray}
j_{(IV)}&=&\epsilon_{abc}\epsilon_{dec}(q_{a}^{T}C\gamma_{5}\gamma_{\mu}s_{b})(\bar{q}_{d}\gamma^{\mu}\gamma_{5}C\bar{Q}_{e}^{T})\nonumber
\end{eqnarray}
for the vector-vector case.
Here $q$ denotes the light $u$ or $d$ quark, $Q$ is the heavy flavor charm quark, and the subscripts $a$,
$b$, $c$, $d$, and $e$ indicate color indices.

\subsection{tetraquark state QCD sum rules}
The two-point correlator
\begin{eqnarray}
\Pi_{i}(q^{2})=i\int
d^{4}x\mbox{e}^{iq.x}\langle0|T[j_{(i)}(x)j_{(i)}^{\dag}(0)]|0\rangle,~~~(i=I, II, III, \mbox{or}~IV)
\end{eqnarray}
can be used to derive QCD sum rules.

Phenomenologically, the correlator can be written as
\begin{eqnarray}\label{ph}
\Pi_{i}(q^{2})=\frac{\lambda_{H}^{2}}{M_{H}^{2}-q^{2}}+\frac{1}{\pi}\int_{s_{0}}
^{\infty}\frac{\mbox{Im}\big[\Pi_{i}^{\mbox{phen}}(s)\big]}{s-q^{2}}ds+...,
\end{eqnarray}
where $s_0$ is the continuum threshold,
$M_{H}$ denotes the hadron's mass, and $\lambda_{H}$ shows
the coupling of the current to the hadron $\langle0|j|H\rangle=\lambda_{H}$.

Theoretically, the correlator can be expressed as
\begin{eqnarray}\label{ope}
\Pi_{i}(q^{2})=\int_{(m_{c}+m_{s})^{2}}^{\infty}\frac{\rho_{i}(s)}{s-q^{2}}ds+\Pi_{i}^{\mbox{cond}}(q^{2}),
\end{eqnarray}
where $m_{c}$ is the mass of charm quark, $m_{s}$ is the mass of strange quark,
and the spectral density $\rho_{i}(s)=\frac{1}{\pi}\mbox{Im}\big[\Pi_{i}(s)\big]$.

After matching Eqs. (\ref{ph}) and (\ref{ope}), assuming quark-hadron duality, and
making a Borel transform $\hat{B}$, the sum rule can be
\begin{eqnarray}\label{sumrule1}
\lambda_{H}^{2}e^{-M_{H}^{2}/M^{2}}&=&\int_{(m_{c}+m_{s})^{2}}^{s_{0}}\rho_{i}(s)e^{-s/M^{2}}ds+\hat{B}\Pi_{i}^{\mbox{cond}},
\end{eqnarray}
with $M^2$ the Borel parameter.

Taking
the derivative of Eq. (\ref{sumrule1}) with respect to $-\frac{1}{M^2}$ and then dividing
by Eq. (\ref{sumrule1}) itself, one can arrive at the hadron's mass sum rule
\begin{eqnarray}\label{sum rule 1}
M_{H}&=&\sqrt{\bigg\{\int_{(m_{c}+m_{s})^{2}}^{s_{0}}\rho_{i}(s)s
e^{-s/M^{2}}ds+\frac{d\big(\hat{B}\Pi_{i}^{\mbox{cond}}\big)}{d(-\frac{1}{M^{2}})}\bigg\}/
\bigg\{\int_{(m_{c}+m_{s})^{2}}^{s_{0}}\rho_{i}(s)e^{-s/M^{2}}ds+\hat{B}\Pi_{i}^{\mbox{cond}}\bigg\}}.
\end{eqnarray}

In detail, the
spectral density
\begin{eqnarray}
\rho_{i}(s)=\rho_{i}^{\mbox{pert}}(s)+\rho_{i}^{\langle\bar{q}q\rangle}(s)+\rho_{i}^{\langle
g^{2}G^{2}\rangle}(s)+\rho_{i}^{\langle
g\bar{q}\sigma\cdot G q\rangle}(s)+\rho_{i}^{\langle\bar{q}q\rangle^{2}}(s)+\rho_{i}^{\langle
g^{3}G^{3}\rangle}(s)+\rho_{i}^{\langle\bar{q}q\rangle\langle
g^{2}G^{2}\rangle}(s)\nonumber
\end{eqnarray}
and the term
\begin{eqnarray}
\hat{B}\Pi_{i}^{\mbox{cond}}&=&\hat{B}\Pi_{i}^{\langle\bar{q}q\rangle\langle
g^{2}G^{2}\rangle}+\hat{B}\Pi_{i}^{\langle\bar{q}q\rangle\langle g\bar{q}\sigma\cdot G q\rangle}
+\hat{B}\Pi_{i}^{\langle\bar{q}q\rangle^{3}}+\hat{B}\Pi_{i}^{\langle\bar{q}q\rangle\langle
g^{3}G^{3}\rangle}+\hat{B}\Pi_{i}^{\langle g^{2}G^{2}\rangle\langle g\bar{q}\sigma\cdot G q\rangle}
+\hat{B}\Pi_{i}^{\langle g\bar{q}\sigma\cdot G q\rangle^{2}}\nonumber\\&+&
\hat{B}\Pi_{i}^{\langle\bar{q}q\rangle^{2}\langle g^{2}G^{2}\rangle}
+\hat{B}\Pi_{i}^{\langle\bar{q}q\rangle^{2}\langle g\bar{q}\sigma\cdot G q\rangle}+\hat{B}\Pi_{i}^{\langle\bar{q}q\rangle\langle g^{2}G^{2}\rangle^{2}}+\hat{B}\Pi_{i}^{\langle g^{3}G^{3}\rangle\langle g\bar{q}\sigma\cdot G q\rangle}+\hat{B}\Pi_{i}^{\langle\bar{q}q\rangle^{2}\langle g^{3}G^{3}\rangle}\nonumber\\&+&
\hat{B}\Pi_{i}^{\langle\bar{q}q\rangle\langle g^{2}G^{2}\rangle\langle g\bar{q}\sigma\cdot G q\rangle}\nonumber
\end{eqnarray}
including condensates up to dimension $12$ can be derived with the similar techniques as Refs. e.g. \cite{overview4,Zhang}.
In reality, their concrete expressions for $\rho_{i}(s)$ and $\hat{B}\Pi_{i}^{\mbox{cond}}$ are the same as our previous work \cite{X5568-SR}
other than that $m_{Q}$ should be replaced by the charm quark mass $m_{c}$, which are
not intended to list here for conciseness. Note that in Ref. \cite{X5568-SR}
we have already applied the factorization hypothesis $\langle\bar{q}q\bar{q}q\rangle=\rho\langle\bar{q}q\rangle^{2}$ \cite{overview2,overview4}
and taken the factorization parameter $\rho=1$.

\subsection{numerical analysis with $\rho=1$}
In the first instance, we set $\rho=1$ for the $\langle\bar{q}q\bar{q}q\rangle=\rho\langle\bar{q}q\rangle^{2}$ factorization.
To extract the numerical value of $M_{H}$,
we perform the analysis of sum rule (\ref{sum rule 1})
and take $m_{c}$
as the running charm quark mass $1.27\pm0.03~\mbox{GeV}$
along with other input parameters as
$m_{s}=96^{+8}_{-4}~\mbox{MeV}$,
$\langle\bar{q}q\rangle=-(0.24\pm0.01)^{3}~\mbox{GeV}^{3}$,
$\langle\bar{s}s\rangle=m_{0}^{2}~\langle\bar{q}q\rangle$,
$\langle
g\bar{q}\sigma\cdot G q\rangle=m_{0}^{2}~\langle\bar{q}q\rangle$,
$m_{0}^{2}=0.8\pm0.1~\mbox{GeV}^{2}$, $\langle
g^{2}G^{2}\rangle=0.88\pm0.25~\mbox{GeV}^{4}$, and $\langle
g^{3}G^{3}\rangle=0.58\pm0.18~\mbox{GeV}^{6}$ \cite{svzsum,overview2,PDG}.
As a standard procedure,
both the OPE convergence and the pole dominance should be considered
to find proper work windows for the threshold $\sqrt{s_{0}}$ and the Borel
parameter $M^{2}$:
the lower bound for $M^{2}$ is obtained by analyzing the OPE
convergence; the upper bound is gained by the
consideration that the pole contribution should be larger
than QCD continuum contribution.
Moreover, $\sqrt{s_{0}}$ characterizes the
beginning of continuum states and can not be taken at will.
It is correlated to
the energy of the next excited state
and approximately
taken as $400\sim600~\mbox{MeV}$ above the extracted
mass value $M_{H}$, which is consistent with
the existing QCD sum rule works
on the same tetraquark state (such
as Refs. \cite{tetra0-heavy-limit,tetra1,tetra3}).

Taking the scalar-scalar case as an example,
its different dimension OPE contributions
are compared as a function of $M^2$
in FIG. 1. Graphically,
one can see that there
are three main condensate contributions, i.e. the dimension $3$ two-quark condensate,
the dimension $5$ mixed condensate, and the dimension $6$ four-quark condensate.
These condensates could play an important role on the OPE side.
The direct consequence is that it is of difficulty to choose a so-called ``conventional Borel window"
namely strictly satisfying that the low dimension condensate should be bigger than the high
dimension contribution. Coming to think of it,
these main condensates
could cancel each other out
to some extent. Meanwhile, most
of other high dimension condensates involved are very small,
for which can not radically influence the character of OPE convergence.
All of these factors make that
the perturbative term could play an important role on the total OPE contribution
and the convergence of OPE is still under control
at the relatively low value of $M^{2}$, and
the lower bound of $M^{2}$ is taken as $0.8~\mbox{GeV}^{2}$ for this case.

In the phenomenological side,
a comparison between pole contribution and
continuum contribution of sum rule (\ref{sumrule1}) for the threshold
$\sqrt{s_{0}}=2.8~\mbox{GeV}$ is shown in FIG. 2, which manifests that the relative pole
contribution is about $50\%$ at $M^{2}=1.6~\mbox{GeV}^{2}$
and decreases with $M^{2}$. In a similar way, the upper bounds of Borel parameters are
$M^{2}=1.5~\mbox{GeV}^{2}$ for
$\sqrt{s_0}=2.7~\mbox{GeV}$ and $M^{2}=1.7~\mbox{GeV}^{2}$ for
$\sqrt{s_0}=2.9~\mbox{GeV}$.
Thereby, Borel windows for
the scalar-scalar case are taken as
$0.8\sim1.5~\mbox{GeV}^{2}$ for $\sqrt{s_0}=2.7~\mbox{GeV}$,
$0.8\sim1.6~\mbox{GeV}^{2}$ for
$\sqrt{s_0}=2.8~\mbox{GeV}$, and
$0.8\sim1.7~\mbox{GeV}^{2}$ for $\sqrt{s_0}=2.9~\mbox{GeV}$.
In FIG. 3, the mass value $M_{H}$
as a function of $M^2$ from sum rule (\ref{sum rule 1})
for the scalar-scalar case is
shown and one can visually see that there are indeed
stable Borel plateaus.
In the chosen work windows,
$M_{H}$ is calculated to be $2.37\pm0.33~\mbox{GeV}$.
Furthermore, in view of the uncertainty due to the variation of quark masses and
condensates, we have
$2.37\pm0.33^{+0.17}_{-0.03}~\mbox{GeV}$ (the
first error is resulted from the variation of $\sqrt{s_{0}}$
and $M^{2}$, and the second error reflects the uncertainty rooting in the variation of
QCD parameters) or briefly $2.37^{+0.50}_{-0.36}~\mbox{GeV}$
for the scalar-scalar tetraquark state.

For the axial-axial case, its OPE contribution in sum rule
(\ref{sumrule1}) for $\sqrt{s_{0}}=2.8~\mbox{GeV}$
 is shown in Fig. 4 by comparing various dimension contributions. Similarly, the dimension $3$,
$5$, and $6$ condensates
could cancel each other out
to some extent and most
of other dimension condensates are very small. On the other hand, the phenomenological contribution in sum rule
(\ref{sumrule1}) for $\sqrt{s_{0}}=2.8~\mbox{GeV}$ is pictured in Fig. 5.
Eventually, work windows for
the axial-axial case are chosen as
$0.9\sim1.5~\mbox{GeV}^{2}$ for
$\sqrt{s_0}=2.7~\mbox{GeV}$, $0.9\sim1.6~\mbox{GeV}^{2}$ for
$\sqrt{s_0}=2.8~\mbox{GeV}$, and $0.9\sim1.7~\mbox{GeV}^{2}$ for
$\sqrt{s_0}=2.9~\mbox{GeV}$.
The corresponding Borel curves for the axial-axial case are
displayed in FIG. 6 and its mass
is evaluated to be
$2.51\pm0.41~\mbox{GeV}$ in the chosen work windows.
With an eye to the uncertainty from the variation of quark masses and
condensates, for the axial-axial tetraquark state we achieve
$2.51\pm0.41^{+0.20}_{-0.02}~\mbox{GeV}$ (the
first error reflects the uncertainty from the variation of $\sqrt{s_{0}}$
and $M^{2}$, and the second error roots in the variation of
QCD parameters) or shortly $2.51^{+0.61}_{-0.43}~\mbox{GeV}$.

For the pseudoscalar-pseudoscalar case, its various dimension OPE contribution in sum rule
(\ref{sumrule1}) for $\sqrt{s_{0}}=2.8~\mbox{GeV}$ is shown in Fig. 7.
One may see that there are also three main condensates, i.e. the dimension $3$,
$5$, and $6$ condensates.
However, what apparently distinct from the foregoing two cases is that
two main condensates (i.e. the dimension $3$
and $6$ condensates)
have a different sign comparing to the perturbative term,
which leads that the perturbative part and the total OPE even have different signs at length.
The dissatisfactory OPE property causes
that related Borel curves
are rather unstable visually, and it is difficult to find
reasonable work windows for this case.
Accordingly, it is not advisable to continue extracting a numerical result.

For the vector-vector case, its different dimension OPE contribution in sum rule
(\ref{sumrule1}) for $\sqrt{s_{0}}=2.8~\mbox{GeV}$ is shown in Fig. 8.
There appears the analogous problem as the pseudoscalar-pseudoscalar case,
and the most direct consequence
is that corresponding Borel curves
are quite unstable.
Hence it is hard to find appropriate work windows to
grasp an authentic mass value for the vector-vector case.

\subsection{numerical analysis with $\rho=3$ and other discussions}

From the analysis in part C, one could see that high-dimension condensates
have been included in the OPE to improve the $M^{2}$-stability of the sum rules.
It is needed to point out that the included condensates are a part of more general condensate
contributions at a given dimension.
There is another source of uncertainty in
the numerical results.
Namely for the $d=6$ four-quark condensate,
a general factorization $\langle\bar{q}q\bar{q}q\rangle=\rho\langle\bar{q}q\rangle^{2}$
has been hotly discussed in the past \cite{A,B},
where $\rho$ is a constant, which may
be equal to $1$, to $2$, or be even smaller than $1$.
(In particular, in Ref. \cite{C} it is argued that
this factorization may not happen at all.)
Furthermore, there may be a poor quantitative control of
the four-quark condensate
due to the violation of factorization parameter which
could be about $\rho=3\sim4$ \cite{D}.
This feature indicates that
the error quoted in the final result which does not take into account such a violation
may be underestimated. Therefore, it is very necessary
to investigate the effect of the factorization breaking.

Compromisingly, in this part we set $\rho=3$
for the $\langle\bar{q}q\bar{q}q\rangle=\rho\langle\bar{q}q\rangle^{2}$ factorization.
Thus, there will be a factor $3$
for the four-quark condensate $\langle\bar{q}q\rangle^{2}$ and for the related classes of these high-dimension condensates
(i.e. $\langle\bar{q}q\rangle^{3}$, $\langle\bar{q}q\rangle^{2}\langle g\bar{q}\sigma\cdot G q\rangle$, $\langle\bar{q}q\rangle^{2}\langle g^{2}G^{2}\rangle$, and $\langle\bar{q}q\rangle^{2}\langle g^{3}G^{3}\rangle$) in the OPE side. From the similar analysis
process as above, the relevant working windows for the scalar-scalar case are taken as: $M^{2}=0.9\sim1.9~\mbox{GeV}^{2}$ for
$\sqrt{s_{0}}=2.7~\mbox{GeV}$, $M^{2}=0.9\sim2.0~\mbox{GeV}^{2}$ for
$\sqrt{s_{0}}=2.8~\mbox{GeV}$, and $M^{2}=0.9\sim2.1~\mbox{GeV}^{2}$ for
$\sqrt{s_{0}}=2.9~\mbox{GeV}$. The Borel curves for this case are
shown in FIG. 9 and in the chosen windows its mass
is computed to be
$2.23\pm0.18~\mbox{GeV}$. Considering uncertainty due to the variation of quark masses and
condensates, one can achieve the final result $2.23^{+0.78}_{-0.24}~\mbox{GeV}$.

Similarly, with $\rho=3$
the working windows for the axial-axial case are taken as: $M^{2}=0.9\sim1.8~\mbox{GeV}^{2}$ for
$\sqrt{s_{0}}=2.7~\mbox{GeV}$, $M^{2}=0.9\sim1.9~\mbox{GeV}^{2}$ for
$\sqrt{s_{0}}=2.8~\mbox{GeV}$, and $M^{2}=0.9\sim2.0~\mbox{GeV}^{2}$ for
$\sqrt{s_{0}}=2.9~\mbox{GeV}$. The corresponding Borel curves are
shown in FIG. 10 and its mass
is evaluated to be
$2.52\pm0.47~\mbox{GeV}$ in the work windows. With a view to uncertainty due to the variation of quark masses and
condensates, we have the eventual result $2.52^{+0.76}_{-0.52}~\mbox{GeV}$.

By this time, note that the $m_{c}$ value
is taken as the running charm quark mass $1.27\pm0.03~\mbox{GeV}$,
which is often used in the existing literature. Without any
evaluation of the perturbation theory radiative corrections, one can equally use
the pole mass $M_{c}\doteq1.4\sim1.5~\mbox{GeV}$ \cite{D}.
One should take into
account a such ambiguity of the charm quark mass definition
to clarify the effects on numerical results for the choice of mass (running or pole).
After setting $\rho=3$, replacing the charm running mass
by the pole mass, and carrying on the same analysis process
as above, one
can obtain the mass ranges $2.11\sim3.16~\mbox{GeV}$ for the scalar-scalar configuration
and $2.11\sim4.31~\mbox{GeV}$ for the axial-axial case.

Additionally, the way to construct a scalar out of two axial-vector
currents or two vector currents could be not unique.
In general, when one combines two spin $1$ currents one may obtain
states with spin = $0$, $1$ and $2$. To be sure that one is dealing with
a scalar, it is needed to project the combination of the currents into
the spin $0$ channel, which can be done with the help of the
projection operators \cite{E}. Since the direct contraction used here
contains the overlap with the rigorous projection, the results found
in this work can be close enough to the projection disposal.
Certainly, one should note that on the final results there may exist
the source of uncertainty from different treatments of currents.

\begin{figure}[htb!]
\centerline{\epsfysize=5.80truecm\epsfbox{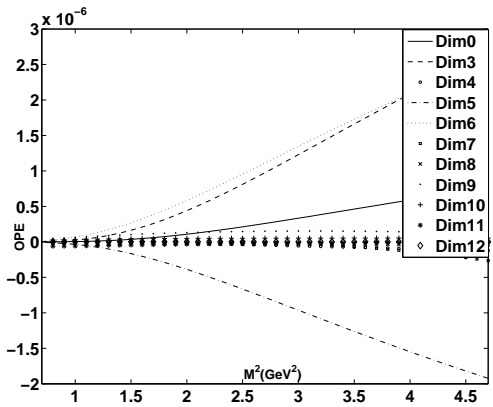}}
\caption{The various dimension OPE contribution as a function of $M^2$ in sum rule
(\ref{sumrule1}) for $\sqrt{s_{0}}=2.8~\mbox{GeV}$ for the scalar-scalar case with $\rho=1$.}
\end{figure}

\begin{figure}
\centerline{\epsfysize=5.80truecm\epsfbox{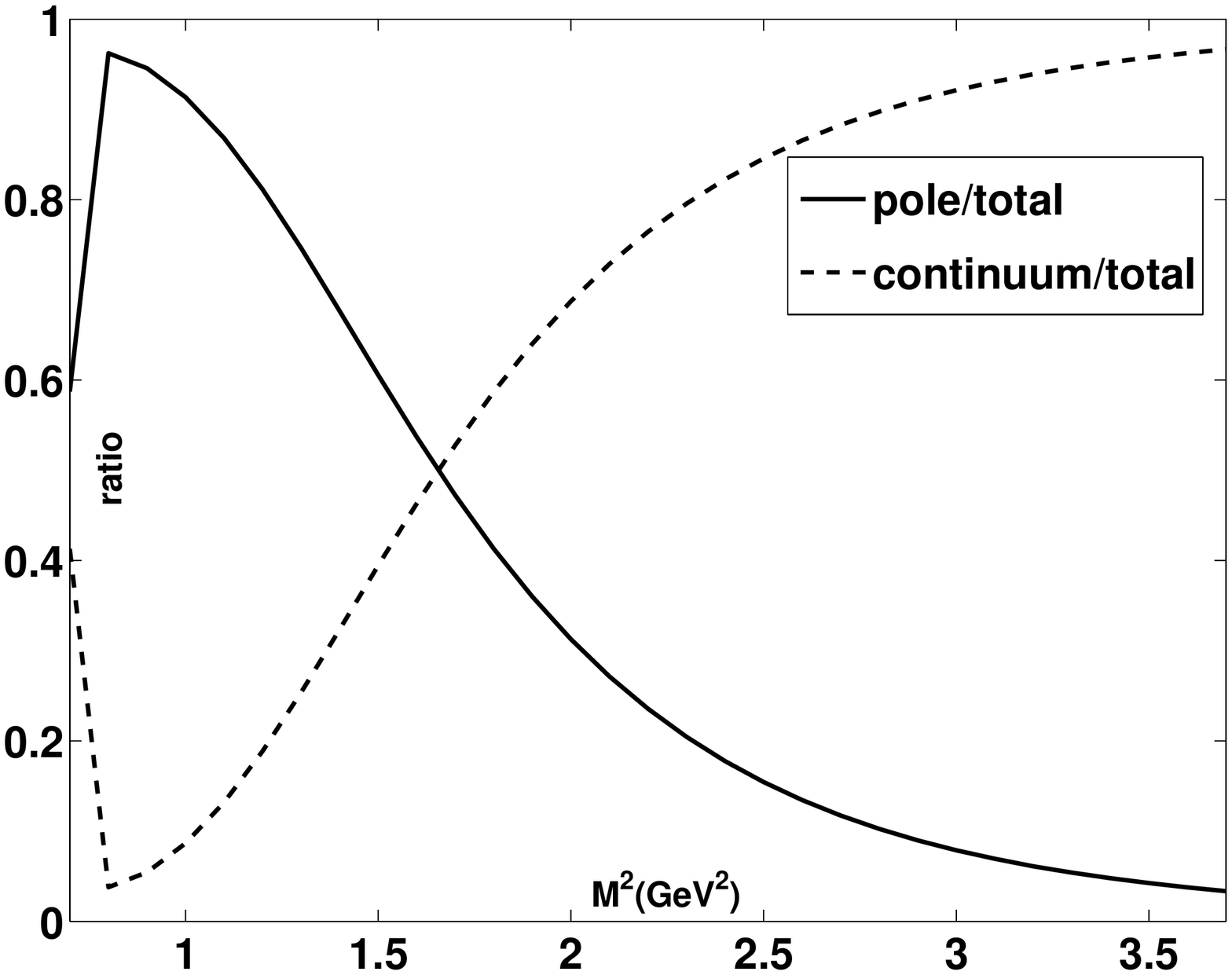}}
\caption{The phenomenological contribution in sum rule
(\ref{sumrule1}) for $\sqrt{s_{0}}=2.8~\mbox{GeV}$ for the scalar-scalar case with $\rho=1$.
The solid line is the relative pole contribution (the pole
contribution divided by the total, pole plus continuum contribution)
as a function of $M^2$ and the dashed line is the relative continuum
contribution.}
\end{figure}

\begin{figure}
\centerline{\epsfysize=5.80truecm
\epsfbox{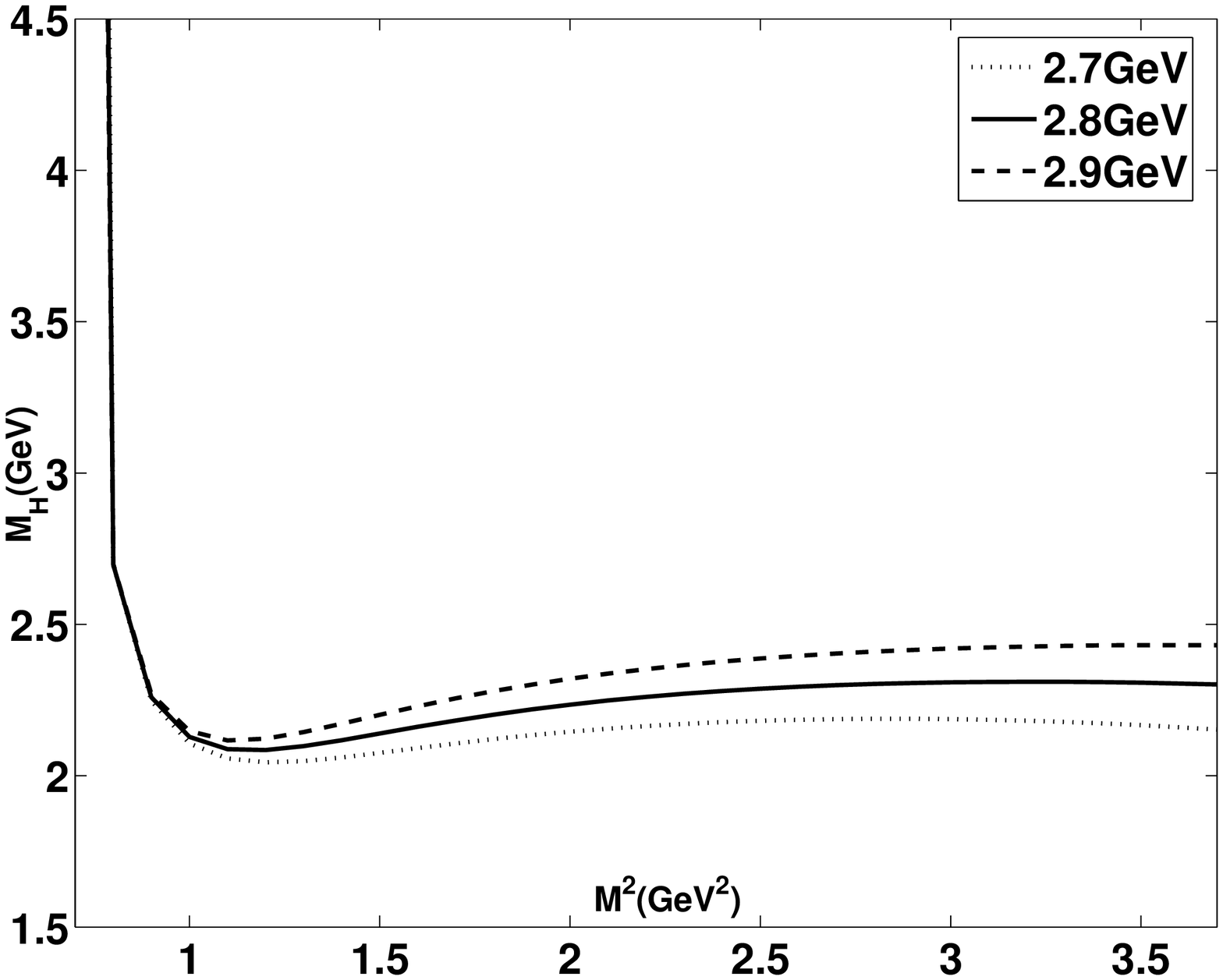}}\caption{
The mass of $0^{+}$ tetraquark state with the scalar-scalar configuration as
a function of $M^2$ from sum rule (\ref{sum rule 1}) with $\rho=1$. The continuum
thresholds are taken as $\sqrt{s_0}=2.7\sim2.9~\mbox{GeV}$. The
ranges of $M^{2}$ are $0.8\sim1.5~\mbox{GeV}^{2}$ for
$\sqrt{s_0}=2.7~\mbox{GeV}$, $0.8\sim1.6~\mbox{GeV}^{2}$ for
$\sqrt{s_0}=2.8~\mbox{GeV}$, and $0.8\sim1.7~\mbox{GeV}^{2}$ for
$\sqrt{s_0}=2.9~\mbox{GeV}$.}
\end{figure}

\begin{figure}
\centerline{\epsfysize=5.80truecm\epsfbox{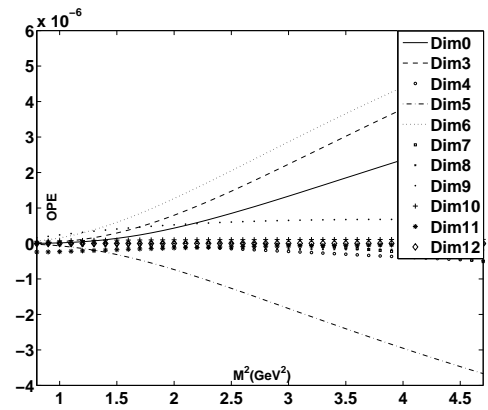}}
\caption{The various dimension OPE contribution as a function of $M^2$ in sum rule
(\ref{sumrule1}) for $\sqrt{s_{0}}=2.8~\mbox{GeV}$ for the axial-axial case with $\rho=1$.}
\end{figure}

\begin{figure}
\centerline{\epsfysize=5.80truecm\epsfbox{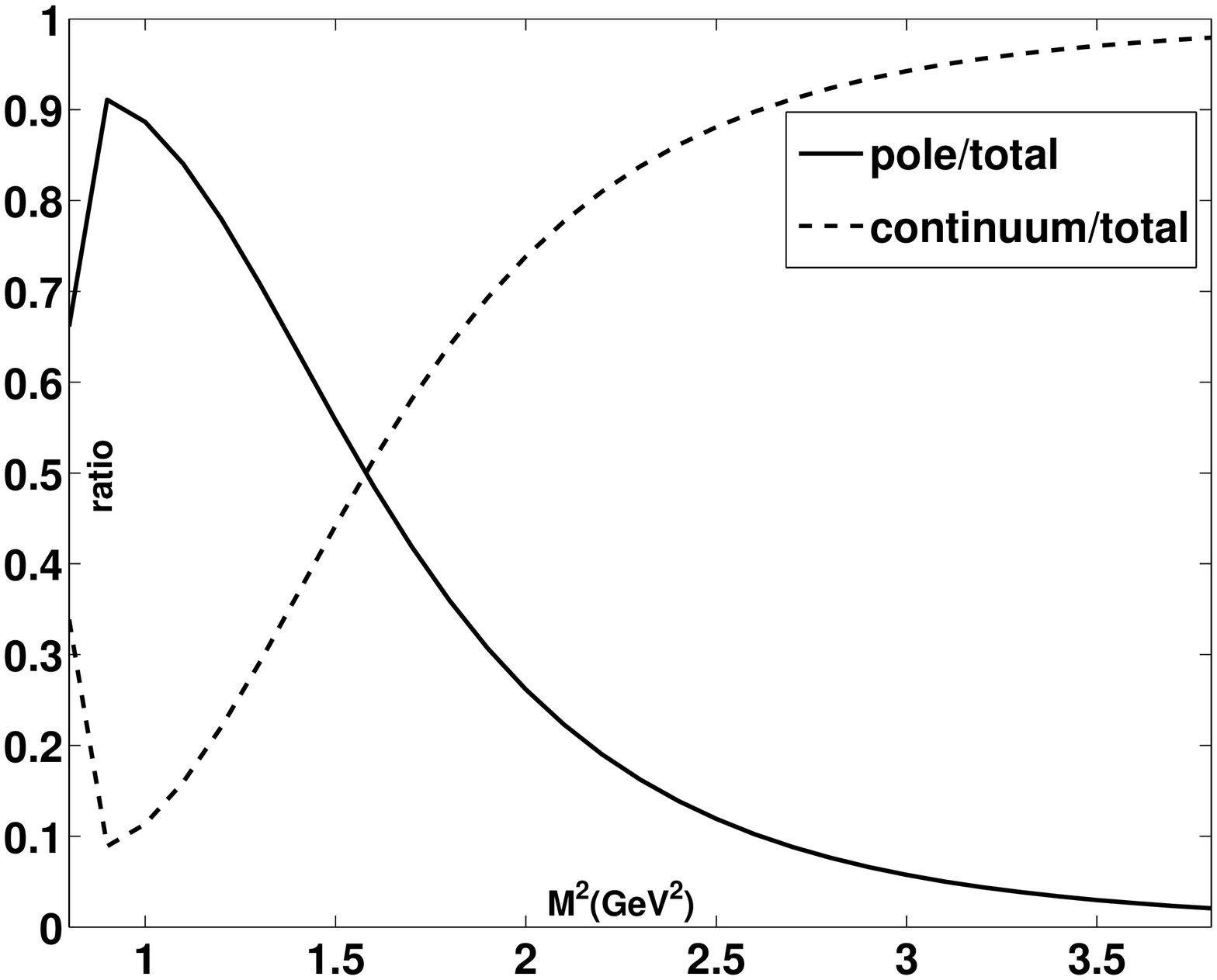}}
\caption{The phenomenological contribution in sum rule
(\ref{sumrule1}) for $\sqrt{s_{0}}=2.8~\mbox{GeV}$ for the axial-axial case with $\rho=1$.
The solid line is the relative pole contribution (the pole
contribution divided by the total, pole plus continuum contribution)
as a function of $M^2$ and the dashed line is the relative continuum
contribution.}
\end{figure}

\begin{figure}
\centerline{\epsfysize=5.80truecm
\epsfbox{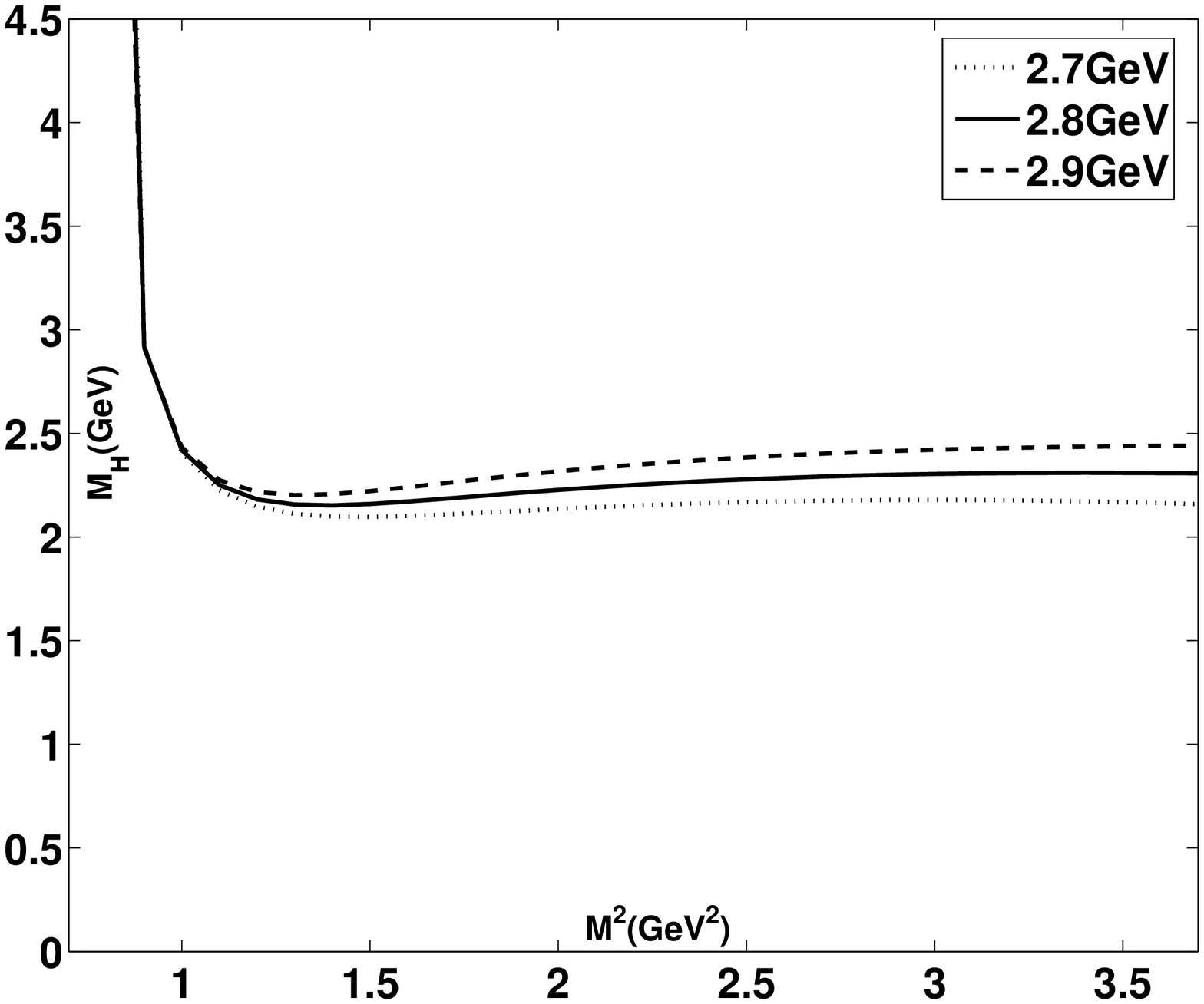}}\caption{
The mass of $0^{+}$ tetraquark state
with the axial-axial configuration as
a function of $M^2$ from sum rule (\ref{sum rule 1}) with $\rho=1$. The continuum
thresholds are taken as $\sqrt{s_0}=2.7\sim2.9~\mbox{GeV}$. The
ranges of $M^{2}$ are $0.9\sim1.5~\mbox{GeV}^{2}$ for
$\sqrt{s_0}=2.7~\mbox{GeV}$, $0.9\sim1.6~\mbox{GeV}^{2}$ for
$\sqrt{s_0}=2.8~\mbox{GeV}$, and $0.9\sim1.7~\mbox{GeV}^{2}$ for
$\sqrt{s_0}=2.9~\mbox{GeV}$.}
\end{figure}

\begin{figure}
\centerline{\epsfysize=5.80truecm\epsfbox{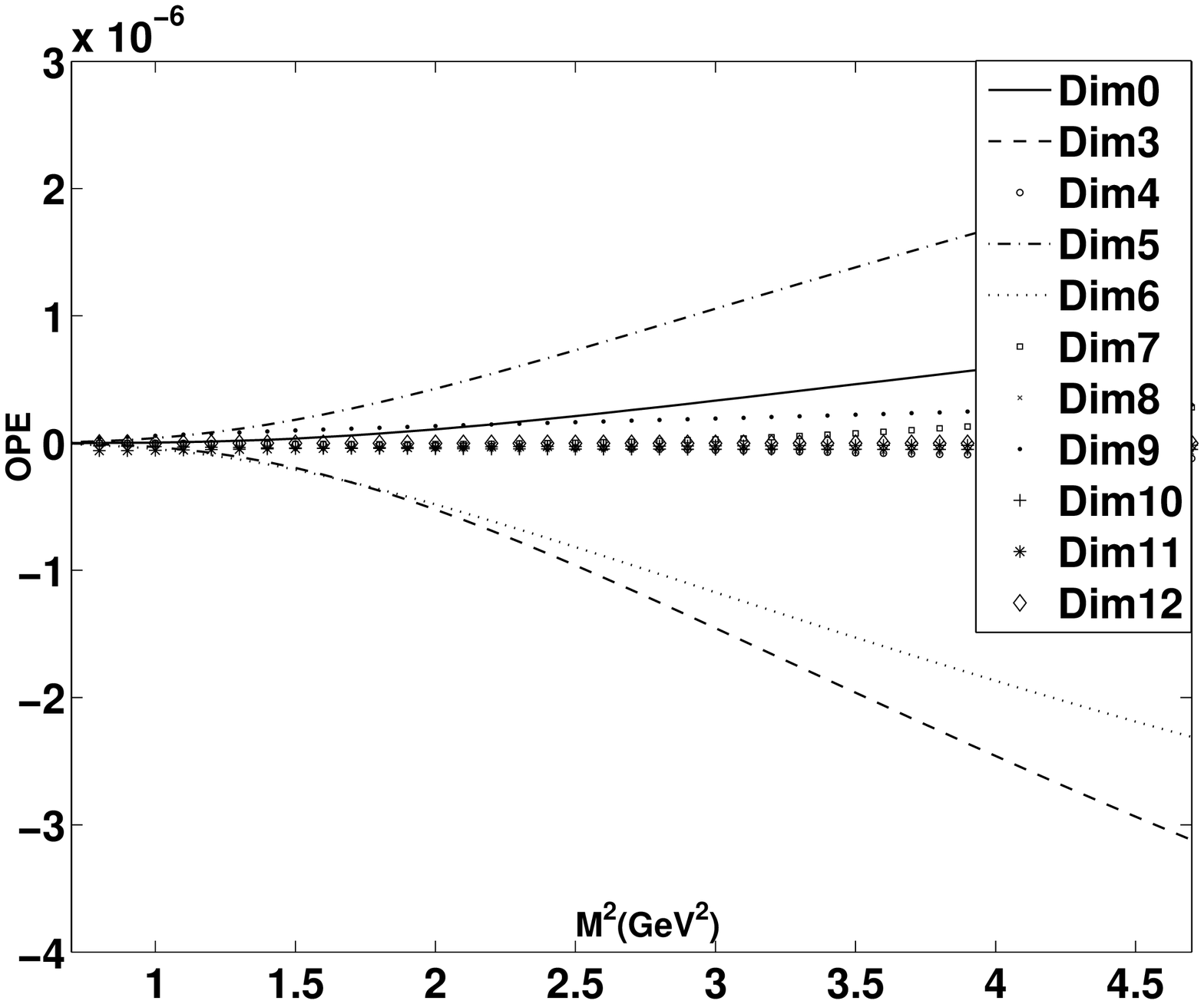}}
\caption{The various dimension OPE contribution as a function of $M^2$ in sum rule
(\ref{sumrule1}) for $\sqrt{s_{0}}=2.8~\mbox{GeV}$ for the pseudoscalar-pseudoscalar case with $\rho=1$.}
\end{figure}

\begin{figure}
\centerline{\epsfysize=5.80truecm\epsfbox{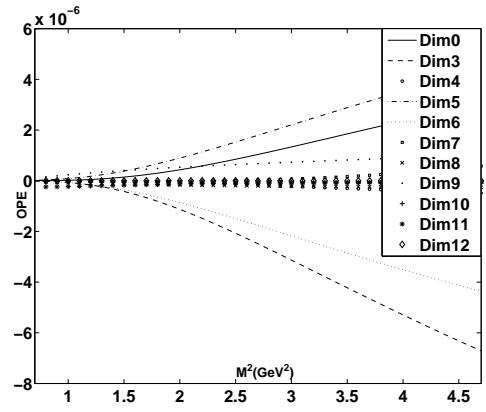}}
\caption{The various dimension OPE contribution as a function of $M^2$ in sum rule
(\ref{sumrule1}) for $\sqrt{s_{0}}=2.8~\mbox{GeV}$ for the vector-vector case with $\rho=1$.}
\end{figure}

\begin{figure}
\centerline{\epsfysize=5.80truecm
\epsfbox{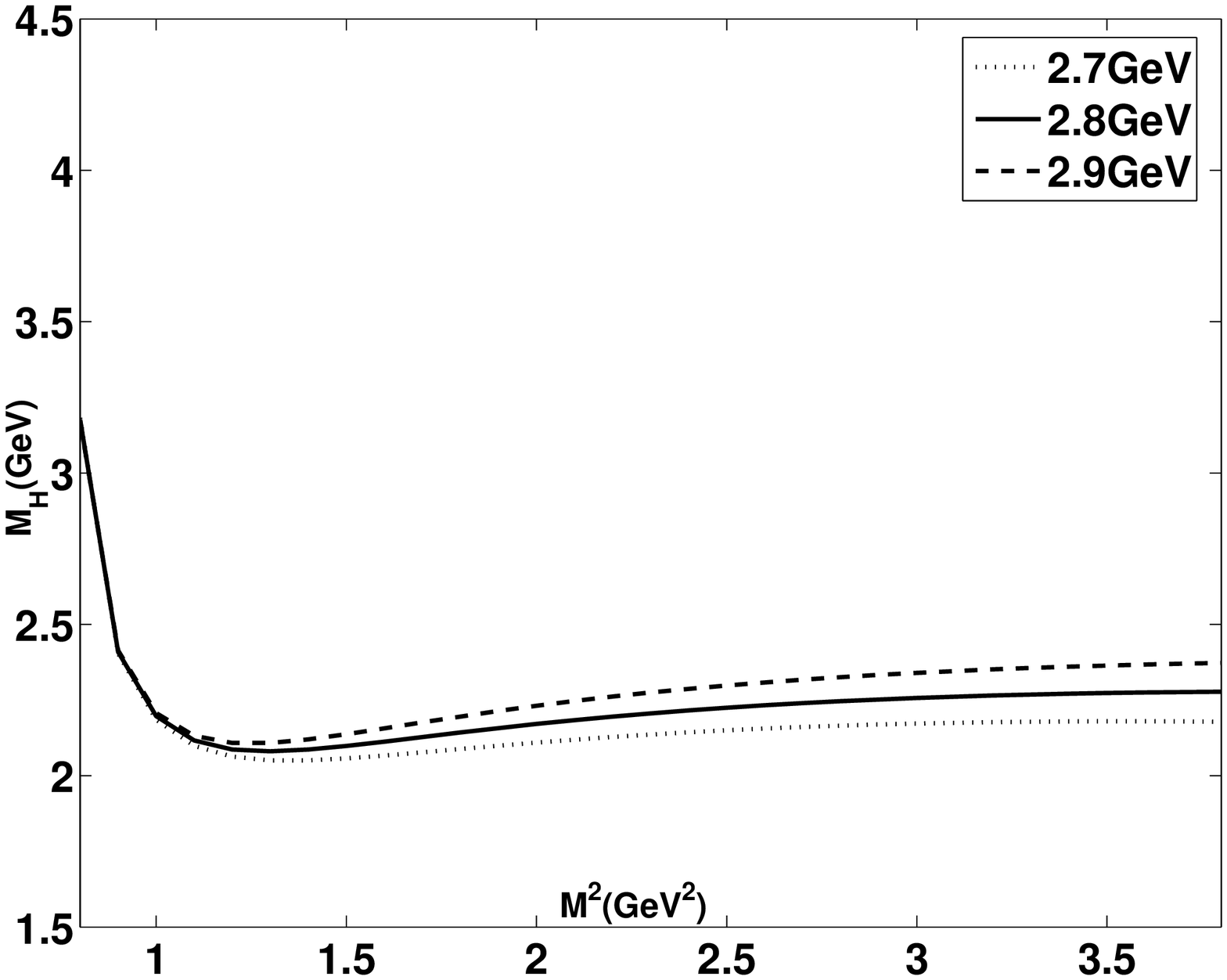}}\caption{
The mass of $0^{+}$ tetraquark state with the scalar-scalar configuration as
a function of $M^2$ from sum rule (\ref{sum rule 1}) with $\rho=3$. The continuum
thresholds are taken as $\sqrt{s_0}=2.7\sim2.9~\mbox{GeV}$. The
ranges of $M^{2}$ are $0.9\sim1.9~\mbox{GeV}^{2}$ for
$\sqrt{s_0}=2.7~\mbox{GeV}$, $0.9\sim2.0~\mbox{GeV}^{2}$ for
$\sqrt{s_0}=2.8~\mbox{GeV}$, and $0.9\sim2.1~\mbox{GeV}^{2}$ for
$\sqrt{s_0}=2.9~\mbox{GeV}$.}
\end{figure}

\begin{figure}
\centerline{\epsfysize=5.80truecm
\epsfbox{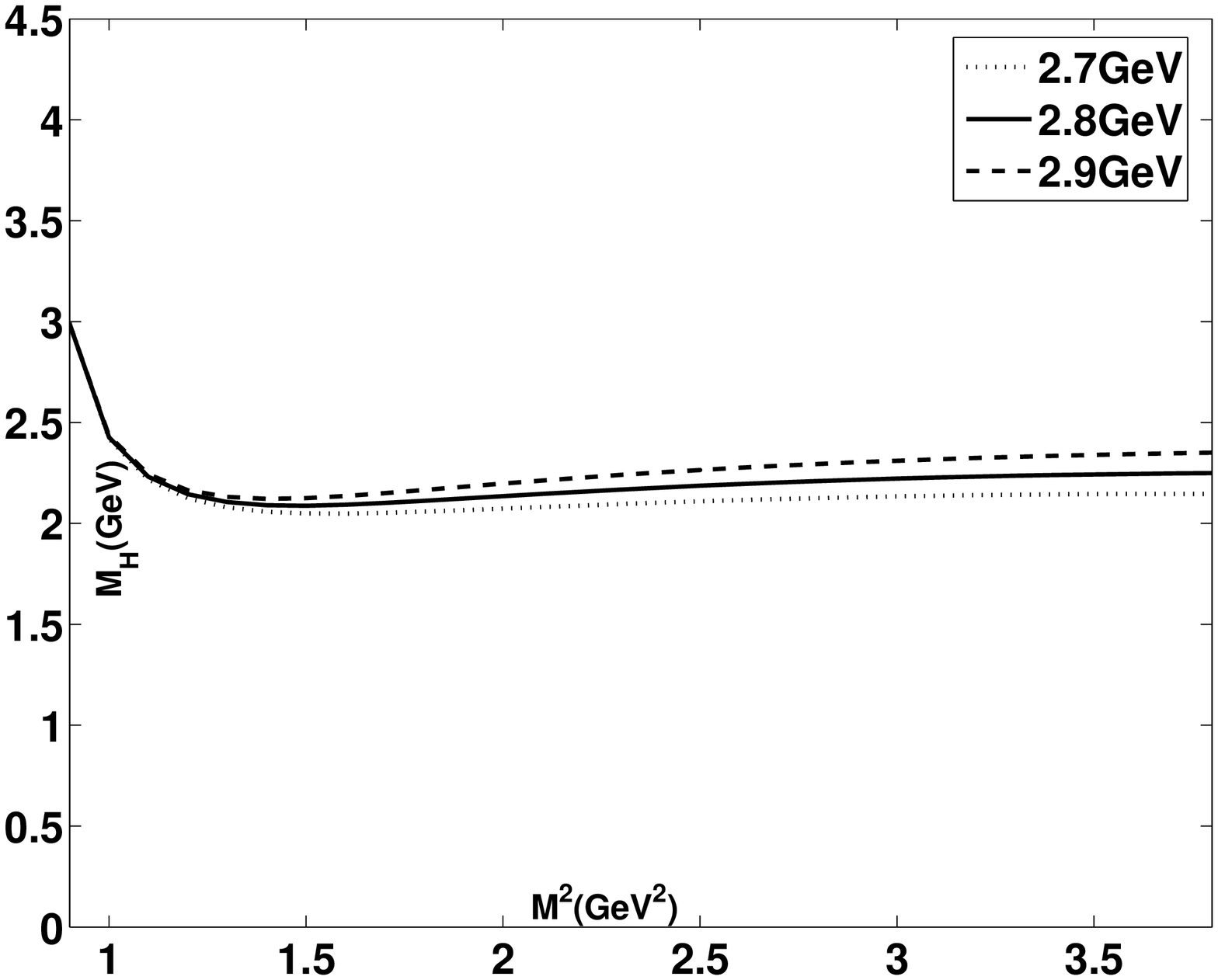}}\caption{
The mass of $0^{+}$ tetraquark state
with the axial-axial configuration as
a function of $M^2$ from sum rule (\ref{sum rule 1}) with $\rho=3$. The continuum
thresholds are taken as $\sqrt{s_0}=2.7\sim2.9~\mbox{GeV}$. The
ranges of $M^{2}$ are $0.9\sim1.8~\mbox{GeV}^{2}$ for
$\sqrt{s_0}=2.7~\mbox{GeV}$, $0.9\sim1.9~\mbox{GeV}^{2}$ for
$\sqrt{s_0}=2.8~\mbox{GeV}$, and $0.9\sim2.0~\mbox{GeV}^{2}$ for
$\sqrt{s_0}=2.9~\mbox{GeV}$.}
\end{figure}

\section{Summary}\label{sec3}
Triggered by the new observation of
$D_{s0}^{*}(2317)$ by BESIII Collaboration, we investigate
that whether
$D_{s0}^{*}(2317)$ could be a $0^{+}$ tetraquark state
employing QCD sum rules.
In order to insure the quality of sum rule analysis, contributions of condensates
up to dimension $12$ have been computed to test the OPE convergence.
We find that some condensates, i.e. the two-quark condensate,
the mixed condensate, and the four-quark condensate are of importance to the OPE side.
Not bad for the scalar-scalar and the axial-axial cases, their main condensates could cancel each other out
to some extent. Most
of other condensates calculated are very small,
which means that they could not radically influence the character of OPE convergence.
All these factors bring that the OPE
convergence for the scalar-scalar and the axial-axial cases is still controllable.

To the end, we gain the following results:
firstly, the final result for the scalar-scalar case is
$2.37^{+0.50}_{-0.36}~\mbox{GeV}$ with the factorization parameter $\rho=1$ (or $2.23^{+0.78}_{-0.24}~\mbox{GeV}$
with $\rho=3$),
which is in good agreement with the experimental value of $D_{s0}^{*}(2317)$.
This result supports that $D_{s0}^{*}(2317)$ could be deciphered as
a $0^{+}$ tetraquark state with the scalar-scalar configuration.
Secondly, the eventual result for the axial-axial case is
$2.51^{+0.61}_{-0.43}~\mbox{GeV}$ with $\rho=1$ (or $2.52^{+0.76}_{-0.52}~\mbox{GeV}$ with $\rho=3$), which is still coincident with the data of $D_{s0}^{*}(2317)$ considering the uncertainty
although its central value is somewhat higher.
In this way, one could not preclude the possibility of
 $D_{s0}^{*}(2317)$ as an axial-axial configuration tetraquark state.
Meanwhile, one should note the
weakness of convergence in the OPE side while presenting these results.
Thirdly, the obtained mass ranges are $2.11\sim3.16~\mbox{GeV}$ for the scalar-scalar configuration
and $2.11\sim4.31~\mbox{GeV}$ for the axial-axial case
while setting $\rho=3$ and taking the charm pole mass, which are both in accord with
the experimental value of $D_{s0}^{*}(2317)$.
Fourthly, the OPE convergence is so unsatisfying for the pseudoscalar-pseudoscalar and the vector-vector cases
that one can not find appropriate work windows to
acquire reliable hadronic information for these two cases.

In the future, with more data accumulated
at BESIII or a fine scan from PANDA \cite{PANDA}, experimental observations
may shed more light on the nature of $D_{s0}^{*}(2317)$.
Besides, one can also expect that the inner structure of $D_{s0}^{*}(2317)$
could be further uncovered by continuously theoretical efforts.

\begin{acknowledgments}
This work was supported by the National
Natural Science Foundation of China under Contract
Nos. 11475258, 11105223, and 11675263, and by the project for excellent youth talents in
NUDT.
\end{acknowledgments}


\end{document}